# A novel hybrid FSO / RF communication system with receive diversity

*Mohammad Ali Amirabadi[1]*
[1] School of Electrical Engineering, Iran University of Science and Technology, Tehran, Iran
✉ E-mail: m_amirabadi@elec.iust.ac.ir

**Abstract:** In mobile communication system, due to the limitations of mobile device such as low power supply as well as small size, most of the processing should be done at the Base Station. Using multi-receive structure at the Base Station really helps better recovery of the original signal by combining different received signals. In this paper, for the first time, receive diversity is used in single-hop hybrid Free Space Optical / Radio Frequency (FSO / RF) communication system. Also it is the first time that a single-hop hybrid FSO / RF system is investigated at saturate atmospheric turbulence regime. For the first time, closed-form expression is derived for Outage Probability of the proposed system and verified through MATLAB simulation. Results indicate a significant improvement in the performance of the proposed structure compared with common FSO and RF systems with receive diversity. Therefore it can be concluded that although the proposed structure requires a complex receiver, but addition of this complexity could significantly reduce processing or power consumption required for performance maintenance of the system.

## 1 INTRODUCTION

Over the last decades, high data rate demands led to more attention to Free Space Optical (FSO) communication system and made it as the main competitor of traditional communication system. This is due to the higher bandwidth and capacity of FSO system compared with Radio Frequency (RF) system. RF system is appropriate in term of cost, but provides lower data rate compared with FSO system. In contrast, FSO system provides both low cost and high data rate [1].

Diversity is a widely used efficient technique for improving performance of communication systems. In mobile communication systems, receive diversity is more practical, because the transmitter (mobile phone) has many limitations and cannot deserve much power consumption or processing complexity. At the receiver side (Base Station), receive diversity technique gathers different copies of the transmitted signal, which are encountered with different fading channels. Combination of these copies could really help better recovery of the transmitted signal [2]. Maximum Ratio Combiner (MRC), Equal Gain Combiner (EGC) and Selection Combiner (SC) are some of the well-known combiners.

Millimetre-wave RF system can achieve data rates equal with FSO system. Weather condition and atmospheric turbulence affect performance of FSO system, and make it un-reliable. However, their impact on FSO and RF links is not the same [3], i.e. in FSO link performance degradation is mostly because of fog and atmospheric turbulence, and heavy rain doesn't affect it. In contrast, RF link is sensitive to heavy rain, and does not care about fog and atmospheric turbulence. Combination of FSO and RF systems is an efficient solution for improving performance of both FSO and RF systems [4].

The so called hybrid FSO / RF systems are available in singe-hop [3, 5-14], dual-hop [15-17], and multi-hop [18-20] structures. Single-hop structure is more preferable for short range communications. In this structure, data is transmitted through two parallel FSO and RF links simultaneously [7,8,13] or by use of a switch [5,6]. Switch technique suffers from continues switches in situations with many changes in weather conditions. In simultaneous transmission this problem no more exists, also it does not require feedback.

In this paper a novel single-hop hybrid FSO / RF system with parallel simultaneous transmission is presents. To the best of author's knowledge it is the first time in a hybrid single-hop FSO / RF system, receive diversity is used. Also it is the first time that a single-hop hybrid FSO / RF system is investigated at saturate atmospheric turbulence regime. Considering RF link in Rayleigh fading and FSO link in Negative Exponential atmospheric turbulence, for the first time, closed form expressions are derived for Outage Probability of the proposed structure. MATLAB simulations verified derived expressions.

In FSO and RF links, ECG and MRC schemes are used, respectively. The main motivation of this kind of selection is that in FSO link, no closed form expression can be derived for probability density function (pdf) and Cumulative Distribution Function (CDF) of MRC scheme in an FSO link with receive diversity in Negative Exponential distribution; also according to discussion in chapter 7 of [21], no closed form expression can be derived for the pdf and CDF of ECG scheme in a RF link with receive diversity in Rayleigh distribution. That is why the combiner of FSO and RF links are not selected the same.

In FSO system, Intensity Modulation and Direct Detection (IM/DD) is used. Combination of FSO and RF links brings advantages of both FSO and RF systems such as reliability, high data rate, security, etc.. On the other hand, Compared with other single-hop FSO / RF systems, the multi-receive structure of the proposed system could really help the receiver to make better decision by combining different received copies of the original signal, which are encountered with different atmospheric turbulence and fading coefficients; therefore, system performance substantially improves.

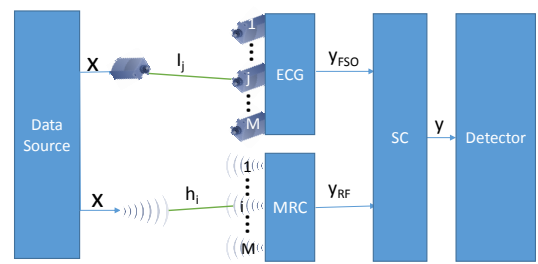

**Fig. 1.** Proposed hybrid FSO/RF system

Rest of the paper is organized as follows: section II describes the system model. Section III derives pdf of the presented structure, section IV derives Outage Probability of the presented system. Section V compares analytical and simulation results. Section VI is conclusion of this study.

## 2  System Model

As shown in Fig. 1, two copies of the original signal are simultaneously transmitted through parallel FSO and RF links. The transmitted signals are multiplied by atmospheric turbulence intensity and fading coefficient, and added by the receiver noise. MRC at the RF receiver, and EGC at the FSO receiver combine the received signals. In conclusion, between output signals of MRC and EGC, signal with maximum Signal to Noise Ratio (SNR) is used for detection.

The SNR at the input of each of MRC and EGC combiners, and thus SNR at the input of SC combiner, is calculated in this section.

### 1.1  FSO link

The pdf of Negative Exponential atmospheric turbulence of is as follows [22]:

$$f_{I_i}(z) = \lambda e^{-\lambda z}, \tag{1}$$

where $I_i$, the atmospheric turbulence intensity of $i-th$; $i = 1,2,\dots,M$ FSO path, has Negative Exponential distribution with $1/\lambda$ mean and $1/\lambda^2$ variance. Assuming $x$ as the transmitted signal, the received signal at the $i-th$ FSO receiver, becomes $y_i = \eta I_i x + e_i$; where $\eta$ is the optical to electrical conversion efficiency and $e_i$, the input noise of $i-th$ FSO receiver, is Additive White Gaussian Noise (AWGN) with zero mean and $\sigma^2$ variance. Output of EGC combiner becomes as follows:

$$y_{FSO} = \sum_{i=1}^{M} y_i = \eta \sum_{i=1}^{M} I_i x + \sum_{i=1}^{M} e_i. \tag{2}$$

Therefore, SNR at the output of EGC is as follows:

$$\gamma_{FSO} = \frac{\eta^2 E[x^2]\left(\sum_{i=1}^{M} I_i\right)^2}{M\sigma^2} = \frac{\bar{\gamma}_1}{M}\left(\sum_{i=1}^{M} I_i\right)^2, \tag{3}$$

where $M$ is number of FSO receiver. $\bar{\gamma}_{FSO} = \eta^2 E[x^2]/\sigma^2$ is average SNR at input of EGC and $E[x^2]$ is transmitted signal energy.

### 1.2  RF link

The pdf of $i-th$; $i = 1,2,\dots,M$ RF path is as follows [21]:

$$f_{|h_i|}(z) = \frac{z}{\sigma_r^2} e^{-\frac{z^2}{2\sigma_r^2}}, \tag{4}$$

where $h_i = r_i e^{j\theta_i}$ is fading coefficient of $i-th$ RF path. Assuming $y_i = h_i x + e_i$ as the received signal at $i-th$ RF receiver, $e_i$ is RF receiver input noise with $\sigma^2$ variance. MRC output is a weighted sum ratio of its input signals. This weight is selected to eliminate the effect of the signal phase and amplifies the signal.

SNR at each branch of MRC is calculated as follows:

$$\gamma_i = \frac{E[|h_i x|^2]}{\sigma^2} = \frac{E[x^2]r_i^2}{\sigma^2} = \bar{\gamma}_i r_i^2, \tag{5}$$

where $\bar{\gamma}_i = E[x^2]/\sigma^2$ is average SNR at the input of $i-th$ RF receiver. SNR at the output of MRC is as follows [21]:

$$\gamma_{RF} = \frac{E[x^2]\sum_{i=1}^{M} r_i^2}{\sigma^2} = \sum_{i=1}^{M} \gamma_i, \tag{6}$$

where $M$ is number of RF receivers.

## 3  Probability density function

In this section closed-form expressions are derived for the pdfs of FSO and RF links.

### 1.3  FSO Link

Moment Generating Function (MGF) of $i-th$ FSO path is as follows:

$$M_{I_i}(s) = \int_0^\infty \lambda e^{-\lambda z} e^{-sz} dz = \frac{\lambda}{s+\lambda}. \tag{7}$$

Assuming independent FSO links achieves [23]:

$$M_{\sum_{i=1}^{M} I_i}(s) = \left(M_{I_i}(s)\right)^M = \left(\frac{\lambda}{s+\lambda}\right)^M. \tag{8}$$

Taking inverse Laplace transform of (8) achieves:

$$f_{\sum_{i=1}^{M} I_i}(z) = L^{-1}\left(M_{\sum_{i=1}^{M} I_i}(s)\right) = \lambda^M \frac{z^{M-1}}{\Gamma(M)} e^{-\lambda z}. \tag{9}$$

Considering [22, Eq. 5.4] and using (9) achieves:

$$f_{\left(\sum_{i=1}^{M} I_i\right)^2}(v) = \frac{\lambda^M}{2\Gamma(M)} v^{\frac{M}{2}-1} e^{-\lambda \sqrt{v}}. \tag{10}$$

Considering [22, Eq. 5.18] and using (10), the pdf of $\gamma_{FSO}$ random variable becomes equal to:

$$f_{\gamma_{FSO}}(\gamma) = \frac{\lambda^M}{2\Gamma(M)\left(\frac{\bar{\gamma}_{FSO}}{M}\right)^{\frac{M}{2}}} \gamma^{\frac{M}{2}-1} e^{-\lambda\sqrt{\frac{\gamma}{\left(\frac{\bar{\gamma}_{FSO}}{M}\right)}}}. \tag{11}$$

### 1.4  RF link

The pdf of $i-th$ RF path is as follows:

$$f_i(\gamma) = \frac{1}{\bar{\gamma}_{RF}} e^{-\frac{\gamma}{\bar{\gamma}_{RF}}}. \tag{12}$$

The MGF of $i-th$ RF path becomes equal to:

$$M_{\gamma_i}(s) = \int_0^\infty \frac{1}{\bar{\gamma}_{RF}} e^{-\frac{\gamma}{\bar{\gamma}_{RF}}} e^{-s\gamma} d\gamma = \frac{\frac{1}{\bar{\gamma}_{RF}}}{s+\frac{1}{\bar{\gamma}_{RF}}}. \tag{13}$$

Assuming independence of RF path fading, achieves [23]:

$$M_{\sum_{i=1}^{M} \gamma_i}(s) = \left(M_{\gamma_i}(s)\right)^M = \frac{\frac{1}{\bar{\gamma}_{RF}^M}}{\left(s+\frac{1}{\bar{\gamma}_{RF}}\right)^M}. \tag{14}$$

Taking inverse Laplace transform of (14), the pdf of $\gamma_{RF}$ random variable becomes equal to:

$$f_{\gamma_{RF}}(\gamma) = L^{-1}\left(M_{\sum_{i=1}^{M}\gamma_i}(s)\right) = \frac{\gamma^{M-1}e^{-\frac{\gamma}{\overline{\gamma}_{RF}}}}{\overline{\gamma}_{RF}^M \Gamma(M)}. \quad (15)$$

## 4 Outage probability

In this section closed-form expressions are derived for individual Outage Probability of FSO and RF links. Multiplying individual Outage Probability of FSO and RF links achieves the Outage Probability of the proposed system. Outage occurs in a system when the input SNR comes down below a threshold, i.e. $\gamma \leq \gamma_{th}$. According to this definition Outage Probability is equal to:

$$P_{out}(\gamma_{th}) = \Pr(\gamma \leq \gamma_{th}). \quad (16)$$

Considering Fig. 1, between output signals of MRC and EGC, signal with maximum SNR is selected by SC, namely:

$$\gamma = \max(\gamma_1, \gamma_2). \quad (17)$$

Therefore, assuming independent FSO and RF links, (16) becomes as follows:

$$P_{out}(\gamma_{th}) = \Pr(\gamma \leq \gamma_{th}) = \Pr(\max(\gamma_{FSO}, \gamma_{RF}) \leq \gamma_{th}) = \Pr(\gamma_{FSO} \leq \gamma_{th}, \gamma_{RF} \leq \gamma_{th}) = P_{out}(\gamma_{FSO} \leq \gamma_{th}) P_{out}(\gamma_{RF} \leq \gamma_{th}). \quad (18)$$

### 1.5 FSO link

By integration of (11) the Outage Probability of FSO link becomes as follows:

$$P_{out_{FSO}}(\gamma_{th}) = \int_0^{\gamma_{th}} \frac{\lambda^M}{2\Gamma(M)\left(\frac{\overline{\gamma}_{FSO}}{M}\right)^{\frac{M}{2}}} \gamma^{\frac{M}{2}-1} e^{-\lambda\sqrt{\frac{\gamma}{(\overline{\gamma}_{FSO}/M)}}} d\gamma. \quad (19)$$

Substituting equivalent Meijer-G form of $e^{-\lambda\sqrt{\gamma/(\overline{\gamma}_{FSO}/M)}}$ as $\frac{1}{\sqrt{\pi}} G_{0,2}^{2,0}\left(\frac{\lambda^2 \gamma}{4\overline{\gamma}_{FSO}/M} \Big| \begin{matrix} - \\ 0, 0.5 \end{matrix}\right)$ [24, Eq.07.34.03.1081.01], the above integral becomes as follows:

$$P_{out_{FSO}}(\gamma_{th}) = \frac{\lambda^M}{2\sqrt{\pi}\Gamma(M)\left(\frac{\overline{\gamma}_{FSO}}{M}\right)^{\frac{M}{2}}} \int_0^{\gamma_{th}} \gamma^{\frac{M}{2}-1} G_{0,2}^{2,0}\left(\frac{\lambda^2 \gamma}{4\overline{\gamma}_{FSO}/M} \Big| \begin{matrix} - \\ 0, 0.5 \end{matrix}\right) d\gamma. \quad (20)$$

Using [24, Eq. 07.34.21.0084.01], Outage Probability of FSO link is equal to:

$$P_{out_{FSO}}(\gamma_{th}) = \frac{\lambda^M}{2\sqrt{\pi}\Gamma(M)\left(\frac{\overline{\gamma}_{FSO}}{M}\right)^{\frac{M}{2}}} \gamma_{th}^{\frac{M}{2}} G_{1,3}^{2,1}\left(\frac{\lambda^2 \gamma_{th}}{4\overline{\gamma}_{FSO}/M} \Big| \begin{matrix} 1-\frac{M}{2} \\ 0, 0.5, -\frac{M}{2} \end{matrix}\right). \quad (21)$$

### 1.6 RF link

By integration of (15), Outage Probability of RF link becomes as follows:

$$P_{out_{RF}}(\gamma_{th}) = \int_0^{\gamma_{th}} \frac{\gamma^{M-1} e^{-\frac{\gamma}{\overline{\gamma}_2}}}{\overline{\gamma}_{RF}^M \Gamma(M)} d\gamma = 1 - e^{-\frac{\gamma_{th}}{\overline{\gamma}_{RF}}} \sum_{k=1}^{M} \frac{\left(\frac{\gamma_{th}}{\overline{\gamma}_{RF}}\right)^{k-1}}{\Gamma(k)}. \quad (22)$$

where the following equation is used to calculate the above integral [24, Eq.01.03.21.0059.01]:

$$\int z^n e^{az} d\gamma = -(-a)^{-n-1} n! \, e^{az} \sum_{k=0}^{n} \frac{(-az)^k}{k!} /; \quad (23)$$

Considering (18), and by multiplying (21) and (22), the Outage Probability of the presented system obtains as follows:

$$P_{out}(\gamma_{th}) = \left(1 - e^{-\frac{\gamma_{th}}{\overline{\gamma}_{RF}}} \sum_{k=1}^{M} \frac{\left(\frac{\gamma_{th}}{\overline{\gamma}_{RF}}\right)^{k-1}}{\Gamma(k)}\right) \times \left(\frac{\lambda^M}{2\sqrt{\pi}\Gamma(M)\left(\frac{\overline{\gamma}_{FSO}}{M}\right)^{\frac{M}{2}}} \gamma_{th}^{\frac{M}{2}} G_{1,3}^{2,1}\left(\frac{\lambda^2 \gamma_{th}}{4\overline{\gamma}_{FSO}/M} \Big| \begin{matrix} 1-\frac{M}{2} \\ 0, 0.5, -\frac{M}{2} \end{matrix}\right)\right). \quad (24)$$

## 5 Simulation Results

This section evaluates performance of presented hybrid FSO / RF system for different variances of Negative Exponential atmospheric turbulence as well as for different number of receivers. The proposed structure is also compared with common FSO and RF systems with receive diversity. FSO and RF receiver inputs are assumed to have equal average SNR ($\overline{\gamma}_{FSO} = \overline{\gamma}_{RF} = \gamma_{avg}$).

In Fig. 2, Outage Probability of the presented hybrid FSO / RF system is plotted in terms of average SNR for different number of receivers for unit variance of Negative Exponential atmospheric turbulence and $\gamma_{th} = 10 dB$. As can be seen at $P_{out} = 10^{-5}$, there is about $8 dB$ difference in $\gamma_{avg}$ between cases of M=1 and M=2, thereby addition of only one receiver significantly decreases power consumption, hence proposed system is suitable for situations in which poser consumption is the major challenge in communication. Power consumption is a major problem especially in saturated atmospheric turbulence regimes, because this kind of turbulence usually occurs at bad weather climates such as seas, in which providing the required power for communication is the major problem, and a system with low power consumption is suitable for this conditions.

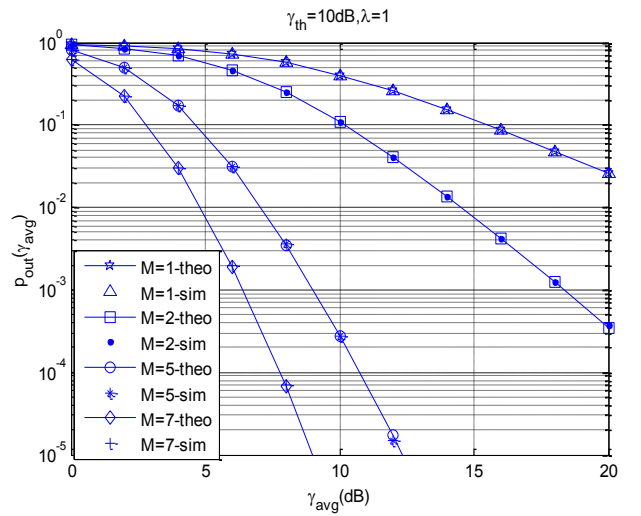

**Fig. 2.** Outage Probability of the presented hybrid FSO / RF system in terms of average SNR for different number of receivers for unit variance of Negative Exponential atmospheric turbulence and $\gamma_{th} = 10 dB$.

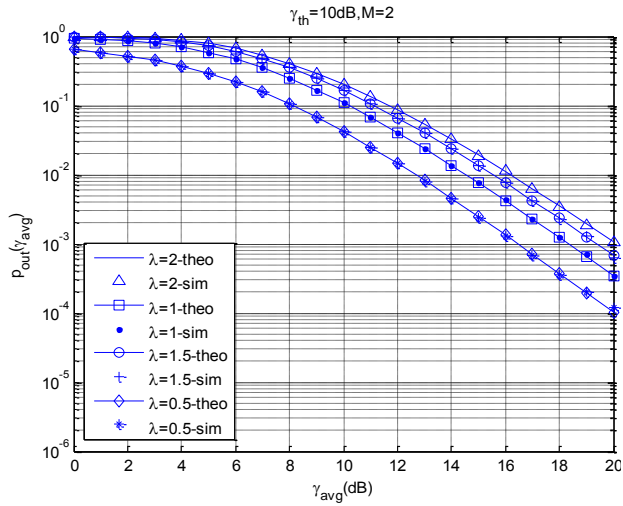

**Fig. 3.** Outage Probability of the presented hybrid FSO / RF system with receive diversity in terms of average SNR, for different atmospheric turbulence variances when number of receiver is M=2 and $\gamma_{th} = 10dB$.

Fig. 3 compares Outage Probability of the presented hybrid FSO / RF system with receive diversity in terms of average SNR, for different atmospheric turbulence variances when number of receiver is M=2 and $\gamma_{th} = 10dB$. As can be seen, at $P_{out} = 10^{-3}$, there is about $5dB$ between two cases of $\lambda = 0.5$ and $\lambda = 1$, hence a small change in atmospheric turbulence, significantly changes system performance.

In this paper, also performance of common FSO and RF systems with receive diversity is compared with proposed hybrid FSO / RF system. In Fig.4, Outage Probability of common FSO and RF systems, compared with presented FSO / RF system is plotted in terms of average SNR. As can be seen, at low $\gamma_{avg}$, Outage Probability of FSO link is less than RF link and almost equal with proposed hybrid FSO / RF links, but by increase of $\gamma_{avg}$, its performance degrades. At all shown $\gamma_{avg}$, presented FSO / RF system performs better than the two others. This is related to Rayleigh and Negative Exponential distributions, because peak of Negative Exponential pdf is at low SNRs while peak of Rayleigh pdf is in higher SNRs it means that loss of Negative Exponential atmospheric turbulence is higher at high SNRs.

# 6 Conclusion

In this paper, a novel model for hybrid FSO / RF system with receive diversity is presented. In which for the first time receive diversity is used in a one-hop hybrid FSO / RF system. FSO link is modeled by Negative Exponential distribution and RF link is modeled by Rayleigh distribution. Outage Probability of the presented system is evaluated in different atmospheric turbulence variances and different number of receivers. Also the proposed system is compared with common FSO and RF systems with receive diversity. The presented system is recommended to use especially in Mediterranean climate where one of the FSO and RF links is always in outage due to heavy rain or dense fog. Although the proposed received diversity has more complexity, but it should be noted that this complexity significantly reduces power consumption and improves performance of the system.

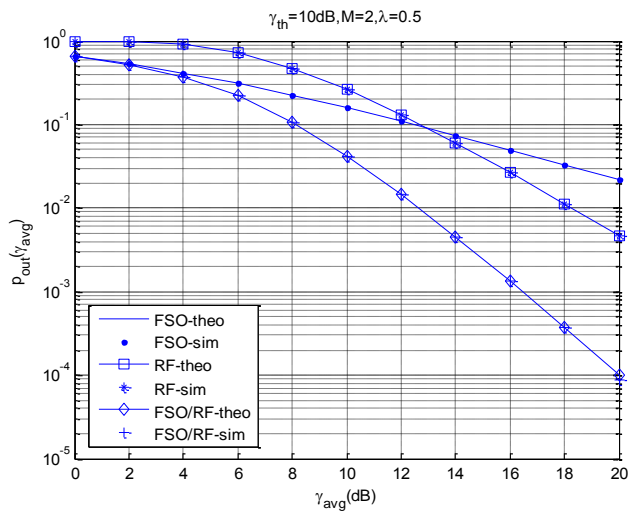

**Fig. 4.** Outage Probability of common FSO and RF systems with receive diversity, compared with proposed hybrid FSO / RF system, in terms of average SNR when $M = 2$, $\lambda = 0.5$ and $\gamma_{th} = 10dB$.